\documentclass{emulateapj}
\usepackage{epsfig}
\usepackage{natbib}
\usepackage{apjfonts}

\shorttitle{Mass Dependent Star Formation Rate Density}
\shortauthors{JUNEAU ET AL.}

\begin{document}

\title{Cosmic Star Formation History
       and its Dependence on Galaxy Stellar Mass}

\author{St\'{e}phanie Juneau\altaffilmark{1,3}, Karl Glazebrook\altaffilmark{2}, 
David Crampton\altaffilmark{3}, Patrick J. McCarthy\altaffilmark{4}, \\
Sandra Savaglio\altaffilmark{2}, Roberto Abraham\altaffilmark{5}, 
Raymond G. Carlberg\altaffilmark{5}, Hsiao-Wen Chen\altaffilmark{6,10}, \\
Damien Le Borgne\altaffilmark{5}, Ronald O. Marzke\altaffilmark{7}, 
Kathy Roth\altaffilmark{8}, Inger J{\o}rgensen\altaffilmark{8}, \\
Isobel Hook\altaffilmark{9}, \& Richard Murowinski\altaffilmark{3}}

\altaffiltext{1}{D\'{e}partement de physique, Universit\'{e} de Montr\'{e}al, 2900,
 boul. \'{E}douard-Montpetit, Montr\'{e}al, QC, Canada H3T 1J4, sjuneau@astro.umontreal.ca}

\altaffiltext{2}{Department of Physics \& Astronomy, Johns Hopkins
University, Baltimore, MD 21218, [kgb; savaglio]@pha.jhu.edu}

\altaffiltext{3}{NRC Herzberg Institute for Astrophysics, 5071 W. Saanich Rd.,
Victoria, BC, Canada, [david.crampton; murowinski]@nrc-cnrc.gc.ca}

\altaffiltext{4}{Carnegie Observatories, 813 Santa Barbara St,
        Pasadena, CA 91101, pmc2@ociw.edu}

\altaffiltext{5}{Department of Astronomy \& Astrophysics, University of Toronto,
        Toronto ON, M5S~3H8 Canada, [abraham; carlberg;
        leborgne]@astro.utoronto.ca}

\altaffiltext{6}{Center for Space Sciences, Massachusetts Institute of
Technology, 70 Vassar St., Bld. 37, Cambridge, MA 02139, hchen@space.mit.edu}

\altaffiltext{7}{Department of Physics and Astronomy, San Francisco State University, San Francisco, CA 94132, marzke@stars.sfsu.edu}

\altaffiltext{8}{Gemini Observatory, 670 North A'ohoku Place, Hilo, HI 97620, [jorgensen; kroth]@gemini.edu}

\altaffiltext{9}{UK Gemini Operations Center, Oxford University, Keble Road, Oxford, OX1 3RH, UK, imh@astro.ox.ac.uk}

\altaffiltext{10}{Hubble Fellow}

\begin{abstract}
We examine the cosmic star formation rate (SFR) and its dependence on galaxy stellar mass over the
 redshift range $0.8 < z < 2$ using data from the Gemini Deep Deep Survey (GDDS). The SFR in the most massive galaxies ($M_{\star} > 10^{10.8} M_{\sun}$) was six times higher at $z = 2$ than it is today. It drops steeply from $z = 2$, reaching the present day value at $z \sim 1$. In contrast, the SFR density of intermediate mass galaxies ($10^{10.2} \le M_{\star} < 10^{10.8} M_{\sun}$) declines more slowly and may peak or plateau at $z \sim 1.5$. We use the characteristic growth time $t_{SFR} \equiv \rho_{M_{\star}}/\rho_{SFR}$ to provide evidence of an
 associated transition in massive galaxies from a burst to a quiescent star formation mode at $z \sim 2$. Intermediate mass systems transit from burst to quiescent mode at $z\sim 1$, while the lowest mass objects undergo bursts throughout our redshift range. Our results show unambiguously that the formation era for galaxies was extended and proceeded from high to low mass systems. The most massive galaxies formed most of their stars in the first $\sim 3$ Gyr of cosmic history. Intermediate mass objects continued to form their dominant stellar mass for an additional $\sim 2$ Gyr, while the lowest mass systems have been forming over the whole cosmic epoch spanned by the GDDS. This view of galaxy formation clearly supports `downsizing' in the SFR where the most massive galaxies form first and galaxy formation proceeds from larger to smaller mass scales. 
\end{abstract}


\section{Introduction}

The evolution of the global star formation rate provides a sensitive probe of galaxy formation and evolution. The earliest determinations of the evolving star formation rate density (SFRD) showed a steep decline from $z \sim 1$ to the present (Lilly et al. 1996; Madau et al. 1996). The behavior of the SFRD at early epochs ($z > 1$) remains uncertain due to variations amongst SFR diagnostics and poorly constrained (yet potentially large) extinction corrections
in the primary rest-frame UV diagnostics (Steidel et al. 1999). Despite these challenges Hopkins (2004) recently compiled results from 33 studies over the range $z = 0$ to $z \approx 6$. The data were used to constrain the luminosity function of star-forming galaxies, and were found to be consistent to within a factor of three over $0 < z < 6$.

Additional insight into star formation histories (SFH) may be gained by the consideration of other physical galaxy properties. Stellar mass is arguably the key parameter. The mass in stars provides a measure of the integral of past galaxy stellar mass assembly, which can be coupled with the instantaneous SFR to give a more complete view of galaxy evolution. 
Mass-based evolution studies are far more deterministic since unlike luminosity, mass evolution is monotonic. While high-luminosity galaxies often evolve into low-luminosity systems, massive galaxies at early epochs must have descendants among the present massive galaxy population. Recent advances in the modeling of multi-color, and particularly near-IR, selected samples lead to fairly robust determinations of stellar masses for galaxies over a wide range of redshifts and luminosities (Brinchmann and Ellis 2000; Fontana et al. 2004; Glazebrook et al. 2004, hereafter Paper III).

Heavens et al. (2004) inferred the star formation history (SFH) of the universe by modeling the spectra of 96,545 local SDSS galaxies. Their results indicate that SFHs vary strongly with {\em present-day\/} stellar mass. Galaxies with $M_{\star} < 10^{10.7} M_{\sun}$ go through the peak of their star formation (SF) at low redshift ($z \leq 0.5$) and $M_{\star} > 10^{11.2} M_{\sun}$ galaxies show an increasing SF activity to $z > 2$. We take advantage of the near-IR selection of the Gemini Deep Deep Survey (GDDS, Abraham et al. 2004, hereafter Paper I) to study directly the evolution of the SFRD as a function of stellar mass at the epoch of observation. Direct measurements of SFRs in the most massive systems at $z > 1$ were made possible only recently, with the progress of spectroscopy of near-IR surveys probing the redshift range $1 < z < 2$. We use ($H_0$, $\Omega_M$, $\Omega_{\Lambda}$) = (70 km s$^{-1}$ Mpc$^{-1}$, 0.3, 0.7) and Vega magnitudes throughout this paper.

\section{Data and Sample Selection}

The GDDS is a spectroscopic survey of an optical and near-IR selected sample targetting massive galaxies at $0.8 < z < 2.0$.  The sample is drawn from the Las Campanas IR imaging survey (McCarthy et al. 2001; Chen et al. 2002), and was designed to select galaxies in the $0.8 < z < 2$ range with an emphasis on the reddest galaxies. Galaxy stellar masses are determined from the mass-to-light ratio $M/L_{K}$ obtained by fitting the $VIK$ photometry with a grid of SED models (Paper III). The sample selection function and weights, details of the observations and catalogs are described in Paper I. The sampling weights are derived as a function of color and magnitude and are used when computing volume-averaged quantities such as the mass density (Paper III) and the SFRD.

The sample selected to compute the SFRD consists of spectra from the GDDS that satisfy: (1) $K<20.6$ (survey limit), (2) redshift confidence level greater than 75\% (${\tt conf} \geq 2$ in Paper I notation) and (3) absence of strong active galactic nuclei activity (${\tt agn} = 0$ in Paper I notation). From the original GDDS sample of 308 spectra, 211 meet the $K$-detection and redshift confidence class criteria. Strong AGN contamination occurs for 1.9\% (4/211) of those objects, bringing the final sample to 207 galaxies.

\section{Method}

Given the redshift range spanned by the GDDS, the available SFR indicators are the [OII]$]\lambda3727$ emission line and the luminosity of the rest-frame UV continuum. For the latter, we chose the absolute rest-frame AB magnitude $M_{2000}$, defined in a synthetic $1900 < \lambda < 2100$ \AA~box filter using an empirical interpolation scheme from the observed $V$ and $I$ magnitudes (Savaglio et al. 2004 (Paper II)). The redshift range is restricted to $z < 1.6$ for SFR([OII]) measurements as the [OII] emission is redshifted out of the optical range at $z > 1.6$. For the SFR($M_{2000}$) measurements, we use the redshift range where the $M_{2000}$ interpolation is 
reliable, i.e. $1.2 < z < 2.0$. Details of the SFR measurements will be given in a forthcoming paper (Juneau et al. in prep.). Briefly, we use the standard SFR($H\alpha$) conversion of Kennicutt (1998) assuming ([OII]$/$H$\alpha)_{obs} = 0.5$ (Glazebrook et al. 1999). Since we use the observed ratio, we need to correct the integrated luminosities at H$\alpha$. We adopt the average extinction of $A_{H\alpha} = 1$ derived from local galaxy samples (Kennicutt 1992).

For SFR measurements based on the rest-frame UV continuum, we apply a dust attenuation correction of the stellar continuum of $A_V = 1$. This is the value typically used in the literature when no direct measure of dust obscuration is available (e.g. Lilly et al. 2003, Sullivan et al. 2000). Following the prescription of Calzetti (2001) this correction is $A_{2000} = 2.2$ at 2000 \AA. 
The mean dust obscuration in galaxies depends on sample selection. $K$-band selection could include more heavily obscured systems. If the mean extinction is greater by 1~mag, the values of SFR (\S 4) and SFRD (\S 5) will shift up by a factor 2.5 whereas $t_{SFR}$ (\S 5) will be lower by the same factor. 

The SFRD is computed with the $1/V_{max}$ method and corrected for both sampling and spectroscopic incompleteness. We define spectroscopic completeness factors that depend on the color and the magnitude in the same fashion as the sampling weights from paper I. We divide the number of spectra with high confidence redshifts (${\tt conf} \geq 2$) by the number of slits in each cell of the color-magnitude plane (see Figure 12 of Paper I). The median spectroscopic completeness of the sample according to this definition is 82\%, consistent with the overall spectroscopic completeness of 79\% for the GDDS.

Throughout this paper, we adopt the Baldry \& Glazebrook (2003) (BG03) Initial Mass Function (IMF), which has a very similar slope to the Salpeter (1955) IMF at high masses and provides a good fit to cosmic and galaxy colors locally. The galaxy masses and SFRs based on BG03 can be used interchangeably with those of Salpeter given the conversion $M(BG03) = 0.55M(SP)$, this ratio is virtually independent of SFH to an accuracy of a few percent.

\section{Star Formation Rates}

The SFRs of the individual galaxies are plotted in Figure~1. The values estimated from the [OII] luminosity (circles) and those obtained with the continuum luminosity at 2000\AA~(triangles) suggest an increase by over one order of magnitude in the upper envelope of the SFR values between $z = 0.8$ and $z = 2$. The color of the plotting symbols is keyed to the stellar mass of the galaxies. The mass bin corresponding to the lowest mass galaxies is $10^{9.0} \le M_{\star} < 10^{10.2} M_{\sun}$ (blue), the intermediate mass is defined as $10^{10.2} \le M_{\star} < 10^{10.8} M_{\sun}$ (green) and the high mass galaxies have $10^{10.8} \le M_{\star} < 10^{11.5} M_{\sun}$ (red). 
Note that $K$-selected samples will miss contributions from low-mass star-forming galaxies fainter than the $K$-limit. (We show later in Figure~2  that our full $K<20.6$ sample underestimates the {\em total\/} SFRD by a factor of 2--3.)

\psfig{file=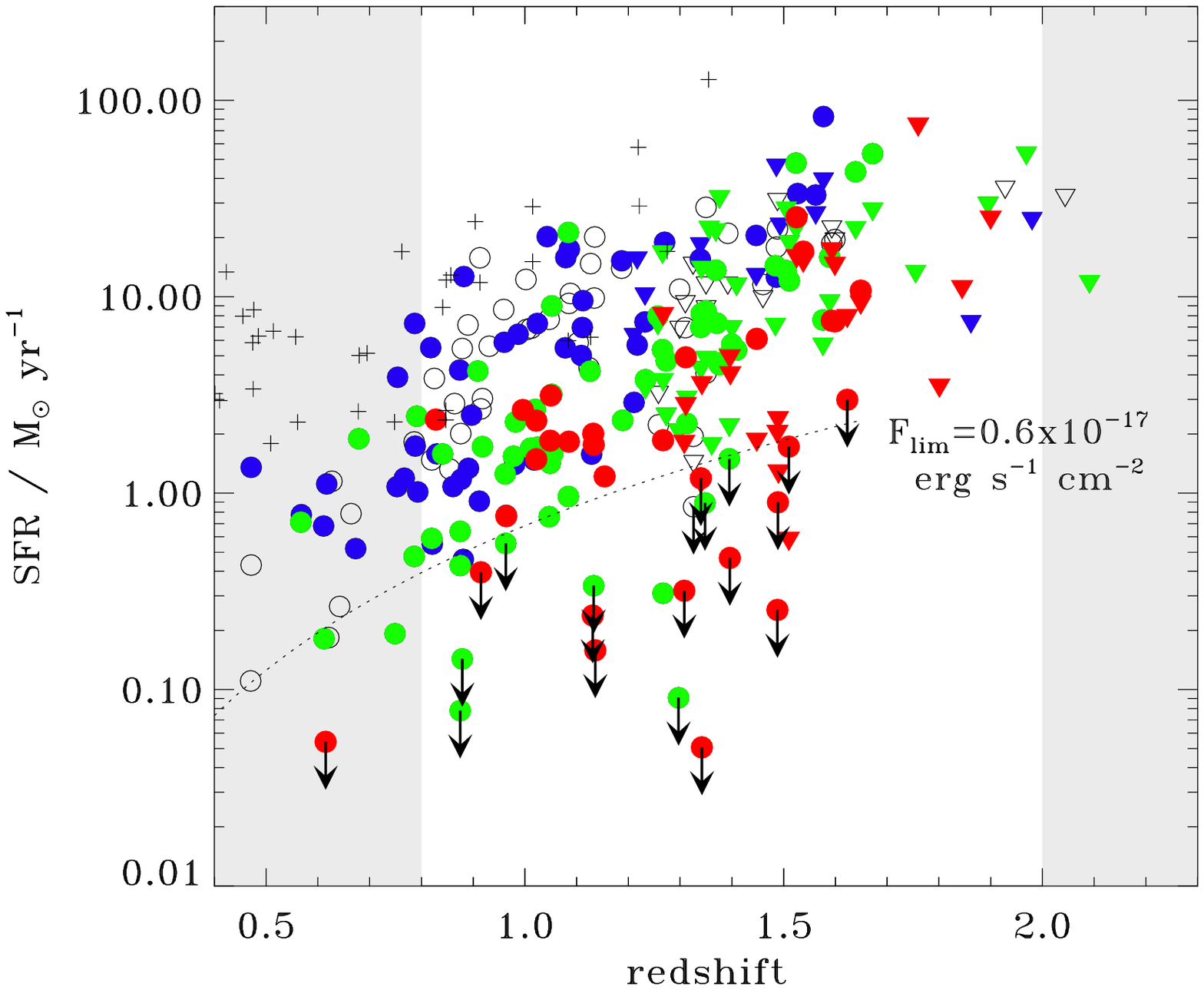,width=8.5cm,angle=0}
\figcaption[]{
SFR derived from L(\rm [OII]) (circles) and from L(2000\AA) (triangles). The values are corrected for obscuration by dust using $A_{H\alpha} = 1$ for emission-line measurements and $A_V = 1$ for UV continuum measurements ($A_{2000} = 2.2$). The open symbols show objects with $K_s > 20.6$ and the symbols color is keyed to the mass bin in which the object belongs: $10^9 M_{\sun} < M_{\star} < 10^{10.2} M_{\sun}$ (blue); $10^{10.2} M_{\sun} < M_{\star} < 10^{10.8} M_{\sun}$ (green) and $10^{10.8} M_{\sun} < M_{\star} < 10^{11.5} M_{\sun}$ (red). The cases where [OII] emission is not formally detected are illustrated with upper limit symbols. The objects with no formal [OII] detection allow us to estimate our [OII] flux sensivity limit $<0.6\times10^{-17}\rm erg s^{-1} cm^{-2}$ (dotted line). A subsample of X-ray star-forming galaxies from HDF-N and Chandra data are shown as a comparison (plus symbols, from Cohen (2003)). The points outside our target redshift range (gray area) are not used in subsequent analysis.}

In nearly 10\% (20/207) of the galaxies, no [OII] emission is detected in our spectra. These systems appear to be primarily massive, quiescent galaxies and their number is uniformly distributed with $z$ (upper limit symbols). The [OII] flux limit varies from one spectrum to another, depending on the redshift and the integration time for each mask. A conservative flux detection limit of $F(\rm[OII])_{lim} = 0.6 \times 10^{-17} erg~s^{-1} cm^{-2}$ is shown as a dotted line in Figure 1. The upper envelope of the [OII]-derived SFRs is representative of star formation in massive galaxies selected at longer wavelength (e.g. $2\mu$m) rather than samples selected in the rest-frame UV. 
To contrast our $K$ selection, Figure 1 also shows a complementary dataset: a subsample of HDF-N galaxies detected by Chandra (shown as plus symbols, and taken from Cohen 2003). Since strong X-ray and [OII] emission is necessary for inclusion in this sample it includes objects with very high SFRs and the sample is not directly comparable. However, it follows the same trend as the GDDS sample. 
Figure~1 demonstrates that the photometric redshift cut used to select galaxies at $0.8 < z < 2.0$ was efficient, as only a handful of objects were observed outside that redshift range (gray area).

The two tracers of SFR were directly compared in the redshift interval in which they overlap: $1.2 < z < 1.6$. 
We found a good linear correlation, with a dispersion $\sim 0.6~dex$ (Juneau et al. in prep.). The scatter might result from the fact that adopting a single value for dust attenuation is not likely to be a realistic representation of a diverse set of galaxies.

\section{Star Formation Rate Density and Characteristic growth Timescale}

The mass selection of the GDDS provides an opportunity to probe cosmic star formation as a function of galaxy stellar mass. In Figure 2 we show the global SFRD for our three mass bins, the total GDDS sample, and the compilation from Hopkins (2004). As in Figure 1, the symbol type is keyed to star formation indicator, circles show [OII] measurements, the triangles represent rest-frame 2000\AA\ based rates. The local SFRDs derived from the SDSS by Brinchmann et al. (2004) in our three mass bins are also shown in Figure 2 (square symbols). These have been transformed to the same IMF and corrected for extinction in the same manner as the GDDS measurements. The comparison with Figure~1 indicates that the increase of SFRD with $z$ is linked with the increase of SFR in individual galaxies.

Paper III shows that for $M_{\star}> 10^{10.8} M_{\sun}$ our sample is mass-complete over our full redshift range. At lower masses, $10^{10.2} \le  M_{\star}< 10^{10.8} M_{\sun}$ we start to become incomplete for $z>1.15$. However the effect is minor because it is the {\em red\/} objects which start to be missed and these contribute {\em least\/} to the SFRD. In this sense, mass-incompleteness also means sensitivity biased towards bluer objects. We can estimate this incompleteness, and calculate a reasonable correction, by bootstrapping from the $z\le 1.15$ complete sample. We calculate that if these objects were placed at $z=1.5$ we would see 90\% of the total SFRD, and at $z=1.8$ we would see 62\%.  Note we are effectively assuming that the high-$z$ sample displays the same range of $M/L_K$ values as the low-$z$ sample. In practice we expect the high-$z$ sample to be bluer, which would make us {\em more} complete. Thus we think our correction is conservative; we extend the upper error bars in Figure~2 to reflect the magnitude of this correction. For $M_{\star}< 10^{10.2} M_{\sun}$ we are incomplete throughout our redshift range and plot
the SFRD values as lower limits.

The key result of this paper is illustrated in Figure 2: the cosmic SFRD is a strong function of galaxian stellar mass. The SFRD in our high mass bin ($M_{\star} > 10^{10.8} M_{\sun}$), while making a minority contribution to the global SFRD, is a factor $\sim 6$ higher at $z = 2$ than it is at present. The SFRD of these massive galaxies has strongly declined since $z = 2$ and reached the present-day level at $z \sim 1$. The SFRD in the intermediate mass bin ($10^{10.2} \le M_{\star} < 10^{10.8} M_{\sun}$) has also steeply declined since $z \sim 1.2$ and appears to have reached a peak or a plateau at a redshift of $\sim 1.5$. The SFRD in the intermediate mass bin then declined to the present day level at $z<1$ and since that time most of the SFRD has been in low mass galaxies. 
Our lowest mass bin ($M_{\star} < 10^{10.2} M_{\sun} $) is strongly mass-incomplete. The comparison with the Hopkins points shows that we miss about half the total SFRD at $z\sim 1$. This would come from low mass galaxies fainter than our survey K limit. 

\psfig{file=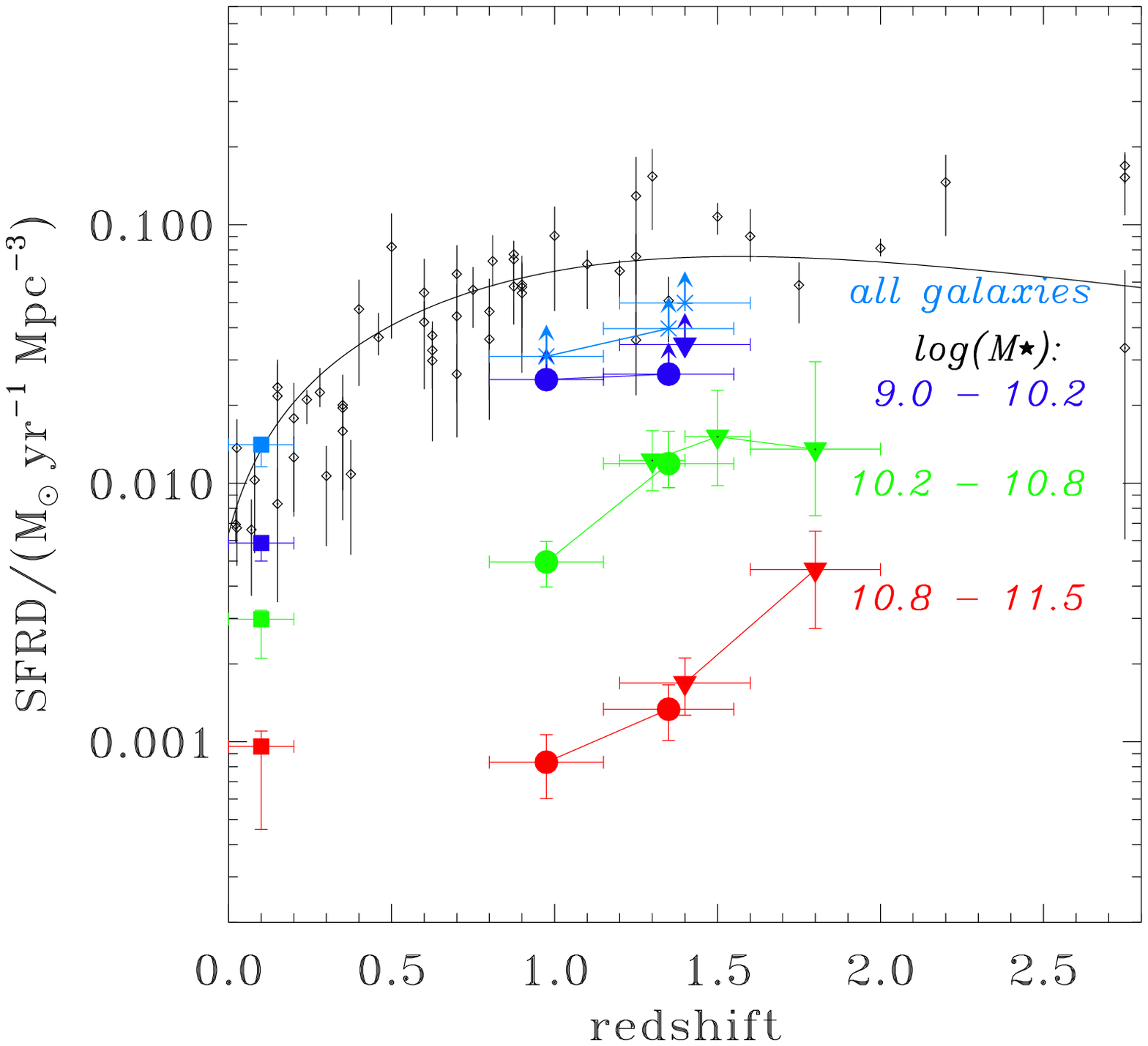,width=8.5cm,angle=0}
\figcaption[]{
SFRD derived from L(\rm [OII]) (circles) and from L(2000\AA) (triangles). The symbols are color-coded with the mass ranges as in Figure~1. 
The error bars in redshift show the width of the redshift bins used. The error bars in SFRD combine shot noise and mass-completeness corrections. Both the sampling and the spectroscopic completeness corrections were applied. The squares are the values found locally by Brinchmann et al. (2004) converted according to our assumed IMF and dust correction. The compilation made by Hopkins (2004), where all the values are converted to a ($\Omega_M=0.3, \Omega_{\Lambda}=0.7, h=0.7$) cosmology, are overplotted with diamonds. The line is the fit derived by Cole et al.(2001) assumming $A_V = 0.6$ (solid line).}

Additional insight into the growth of galaxies can be gained from comparing the stellar mass density to the SFRD over a range of redshifts. The ratio gives a characteristic growth timescale $t_{SFR} = \rho_{M_{\star}} / \rho_{SFR}$ which can be interpreted simply as the time required for the galaxies to assemble their observed stellar mass assuming that their observed SFR stays constant. Results for the GDDS sample are shown in Figure~3.  In this figure we compare $t_{SFR}$ with the Hubble time $t_H(z)$, the age of the universe at a given redshift. If equal, galaxies can form all their observed stars in a Hubble time. At a given redshift, $t_{SFR} > t_H$ suggests that the galaxies are in a declining or quiescent star formation mode at the observed redshift and that the bulk their star formation has occurred in the past. Conversely, $t_{SFR} < t_H$ indicates that the galaxies are going through a burst phase. This past average-to-present SFR 
allows us to investigate the {\em mode\/} of SFR in the SFRD(z) diagram. The highest mass galaxies are in a quiescent mode at low-$z$, transiting to a burst mode at $z \sim 1.8$. The intermediate mass objects make a similar transition at $z \sim 1.1$. The lowest mass systems appear to be observed in a burst mode at all redshifts, though we note our selection is only sensitive to the bluest objects for this mass-incomplete subsample. The transition redshift from burst mode to quiescent mode is a strong function of stellar mass and looks like a time-delayed echo of the corresponding downturn in the SFRD(z) diagram.

\psfig{file=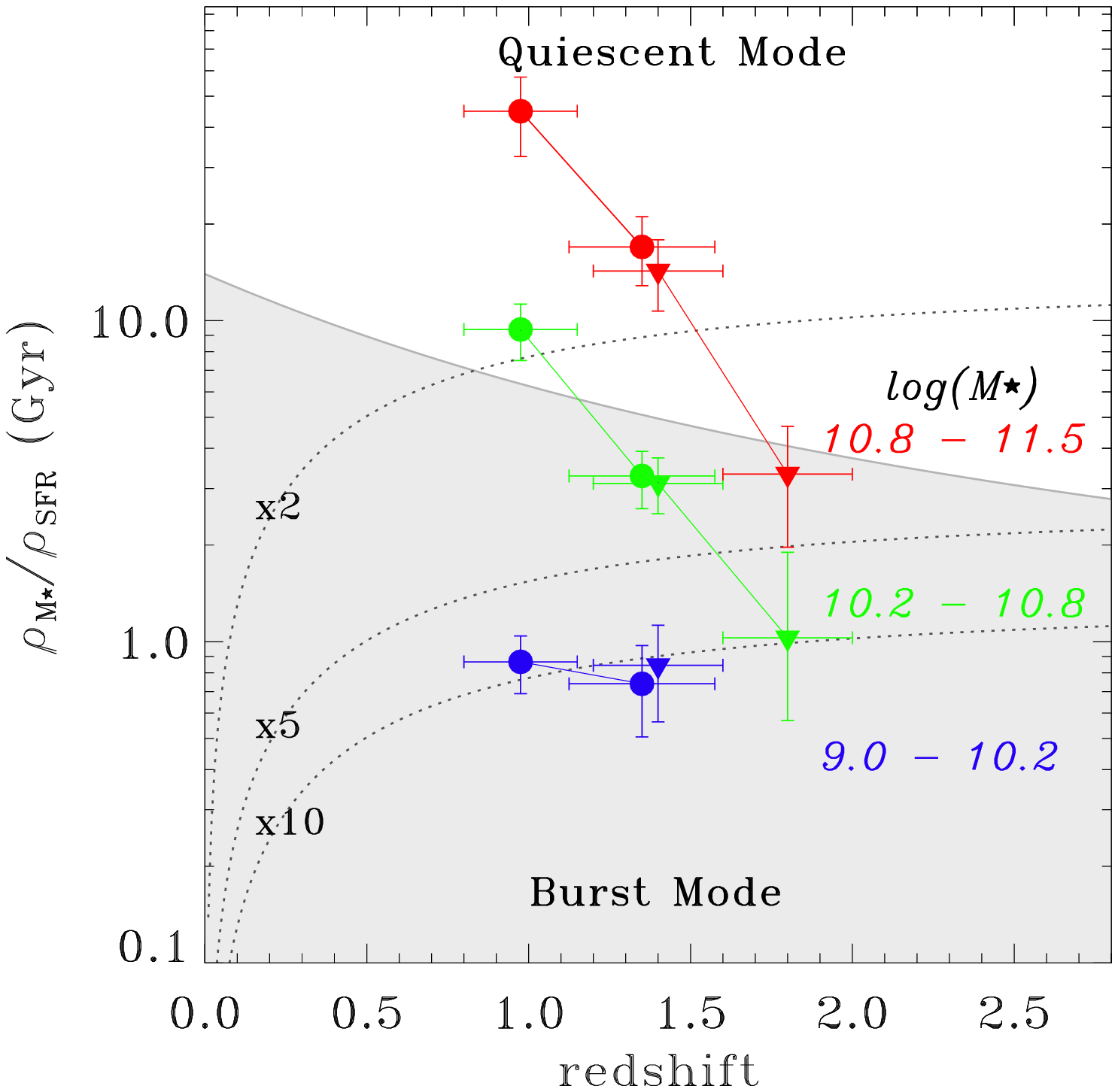,width=8.5cm,angle=0}
\figcaption[]{
Characteristic timescale of stellar mass growth in galaxies. The SFRD is derived from L(\rm [OII]) (circles) and from L(2000\AA) (triangles). The plotting symbols are keyed to the galaxy stellar mass as in Figures 1 \& 2.
The error bars in redshift show the width of the redshift bins used. The error bars in SFRD are statistical. 
The gray line shows the age of the universe in our adopted cosmology. It indicates the transition from quiescent SF mode to burst SF mode (gray area). Along the dotted lines, lookback time to the present allows galaxy stellar mass to increase by a factor of 2, 5 or 10, if the SFR {\em stays} constant, as labeled.}

\section{Discussion and Summary}

Figures 2 and 3 paint a simple picture of galaxy formation in a volume-averaged sense. The upper envelope
of SFR rises with redshift and one sees very clear mass-dependent effects in the SFRD($z$) and $t_{SFR}(z)$ diagrams. The mass-scale of the SFR and of burst activity both decline with time.

The basic conclusion is that the most massive systems formed early and were nearly finished forming their stars  by $z \sim 1.5-2$. This is supported by several other lines of evidence. The reddest galaxies in the $1 < z < 2$ interval appear to have ages $> 1$ Gyr and $z_f > 2$ (McCarthy et al. 2004; Cimatti et al. 2004). These objects are $\sim 10^{11} M_{\sun}$ and suggest that the rising SFRD seen in the high-mass bin of Figure 2 continues to increase, reaching a peak somewhere in the $2 < z < 4$ interval. The most massive {\em local\/} galaxies seen in the SDSS also appear to be dominated by stars with early formation redshifts (Heavens et al. 2004). The Hubble sequence begins to emerge between $1 < z < 2$, and appears to be in place at $z \sim 1$ (Brinchmann \& Ellis 2000).
Disk galaxies that are seen at $z > 2$ are primarily in massive stellar systems (e.g. Labb\'{e} et al 2003; Stockton et al. 2004). As Figure~3 shows, by $z \sim 1$ both the high and intermediate mass populations, those that dominated the elliptical and disk portions of the Hubble sequence, have transitioned to fairly quiescent SF and have formed the bulk of their stars. A puzzle remains in the continued accumulation in the
{\em number} of massive (see Paper III and Fontana et al. 2004) and/or red-sequence (Bell et al. 2004) 
galaxies by a factor of $\sim 2$ over $0<z<1$. This now has to occur without making a large contribution to the global SFRD in the visible and rest-UV, perhaps via mergers that produce either little or heavily obscured star formation.

Our view of galaxy formation reveals a clear `downsizing' of star formation from high-mass to low-mass galaxies with time. This basic picture was first suggested by Cowie et al. (1996). The downsizing picture of galaxy {\em stellar mass\/} assembly is the reverse of the original picture of hierarchical {\em dark matter\/} assembly (Blumenthal et al. 1984) where the large units come last in the sequence. If the hierarchical picture is correct, then star formation must be much more efficient in early times in high-mass systems, as is required (e.g. Paper III) to explain their space densities at high redshift. 

\section{Acknowledgments}

The authors wish to thank J. Brinchmann and the referee for their valuable comments.
Observations were obtained at the Gemini
Observatory, which is operated by  AURA Inc., under a cooperative 
agreement with the NSF on behalf of the Gemini partnership: the NSF (US),
PPARC (UK), NRC (Canada), CONICYT (Chile), ARC
(Australia), CNPq (Brazil) and CONICET (Argentina) and at the Las Campanas
 Observatory of the OCIW. 
KG and SS acknowledge support from the David and Lucille Packard Foundation, 
RGA acknowledges support from the NSERC.

\end{document}